\documentclass[useAMS,usenatbib]{mn2e}

\usepackage{graphicx}
\usepackage{amssymb}
\newcommand{\eqb}{\begin{eqnarray}}
\newcommand{\eqe}{\end{eqnarray}}
\newcommand{\sth}{\sigma_{\rmn T}}
\newcommand{\gmn}{\gamma_{\rm min}}
\newcommand{\gmx}{\gamma_{\rm max}}
\newcommand{\mpr}{m_{\rm p}}
\newcommand{\me}{m_{\rm e}}
\newcommand{\up}{u_{\rm p}}
\newcommand{\ub}{u_{\rm B}}

\newcommand{\lcr}{\ell_{\rm cr}^{\rm inj}}
\newcommand{\lp}{\ell_{\rm p}}
\newcommand{\lgg}{\ell_{\gamma}}
\newcommand{\dmin}{\delta_{\rm min}}
\newcommand{\tcr}{t_{\rm cr}}
\newcommand{\gp}{\gamma_{\rm p}}
\newcommand{\eg}{\epsilon_{\gamma}}
\newcommand{\Bcr}{B_{\rmn cr}}
\newcommand{\Bq}{B_{\rmn q}}
\newcommand{\Beq}{B_{\rm eq}}
\newcommand{\Pjet}{P_{\rm jet}}

\title[On proton synchrotron blazar models: the case of quasar 3C 279]
{On proton synchrotron blazar models: the case of quasar 3C 279}
\author[M. Petropoulou and A. Mastichiadis]{M. Petropoulou (MP) \thanks{E-mail:
maroulaaki@gmail.com } and A. Mastichiadis (AP) \thanks{E-mail:
amastich@phys.uoa.gr}\\
Department of Physics, University of Athens, Panepistimiopolis, GR 15783 Zografos, Greece}
\begin{document}
\date{Received.../Accepted...}

\pagerange{\pageref{firstpage}--\pageref{lastpage}} \pubyear{2012}

\maketitle

\label{firstpage}
\begin{abstract}

In the present work we propose
an innovative estimation method for the minimum
Doppler factor and energy content
of the $\gamma$-ray emitting region of quasar 3C 279, using 
a standard proton synchrotron blazar model
and the principles of automatic photon
quenching. The latter becomes relevant for high enough magnetic fields and
results in spontaneous annihilation of $\gamma$-rays. 
The absorbed energy is then redistributed into electron-positron pairs and soft radiation.
We show that as quenching sets an upper value for the source rest-frame $\gamma$-ray luminosity,
 one has, by neccessity, to resort to Doppler factors that lie above
a certain value in order to explain the TeV observations.
 The existence of this lower limit for the
Doppler factor has also implications on the energetics
of the emitting region. In this aspect, the proposed
method can be regarded
as an extension of the widely used one 
for estimating the equipartition magnetic field using 
radio observations. In our case, the leptonic synchrotron component is
replaced by the proton synchrotron emission and the
radio by the VHE $\gamma$-ray observations.
We show specifically that one can
model the TeV observations by using parameter
values that minimize both the energy density and the
 jet power at the cost of high-values of the Doppler factor. 
On the other hand, the modelling
can also be done by using the minimum possible
Doppler factor; this, however, leads to a particle
dominated region and high jet power
for a wide range of magnetic field values. 
Despite the fact that we have focused on the case of 3C 279,
our analysis can be of relevance to all TeV blazars favoring hadronic
modelling that have, moreover,
simultaneous X-ray observations.

\end{abstract}


\section{Introduction}
Blazars, a subclass of Active Galactic Nuclei, emit 
non-thermal, highly 
variable radiation across the whole electromagnetic spectrum. 
According to the standard scenario,
particles accelerate to relativistic energies   
in the jets of these objects 
which point, within a small angle, towards our direction
and the resulting photon emission is boosted due to relativistic
beaming.

Detailed modelling of observations, especially
in the $\gamma$- and X-ray regimes, makes possible the estimation of the 
physical parameters of the emitting region. Thus quantities
like the source size, the magnetic field strength, the bulk
Lorentz factor and the particle energy density can 
nowadays be routinely calculated.
Furthermore, from the values of these quantities
one could obtain meaningful estimates for the
particle and Poynting fluxes  and use them to make comparisons,
for example, to the Eddington luminosity of the 
source, connecting thus the black hole energetics  with the jet power.

One major uncertainty of the modelling is the nature of the radiating
particles.  
While there seems to be a consensus that the emission from
radio to X-rays comes from the synchrotron radiation of
a population of relativistic
electrons, there are still open questions regarding 
the $\gamma$-ray emission of these objects. Broadly
speaking, the models fall into two  categories,
the leptonic ones (e.g. \citealt*{dermeretal92, maraschietal92, dermerschlick93})
which assume that the same electrons 
which radiate at lower frequencies via synchrotron
produce the $\gamma$-rays by inverse Compton scattering and
the hadronic ones (\citealt*{mannheimbiermann93, mueckeprotheroe01, boettcher09}; \citealt{muecke03})
 which postulate that an extra population of 
relativistic protons produce the high energy
radiation as a result of hadronically induced and electromagnetic processes.

Due to the very different radiating mechanisms involved, the two 
classes of models can result in very different parameters
for the source. Thus, for example, 
typical leptonic synchrotron self-compton models require low magnetic field strengths, 
ranging from $B \simeq 0.01 - 1$ G for high synchrotron-peaked BL Lacs \citep{tavecchio11, murase12} up to $B\simeq 1-10$ G for
 Flat Spectrum Radio Quasars,
and low jet power 
($\simeq 10^{47}$ erg/sec) (e.g. \citealt{celotti08}). On the other hand,
the corresponding 
values for the hadronic models are higher by
at least one order of magnitude (e.g. \citealt{protheroe01, boettcher09}). 
While both can fit 
reasonably well the observations, they both face some problems.  
For instance, it has been argued that one problem with
leptonic models is the high ratio between the required
relativistic electron energy density to the magnetic one,
implying large departures from equipartition. On the other hand, 
the hadronic models imply jet powers which can be 
uncomfortably high, especially when compared to the accretion 
luminosity.

In previous work \citep{petropoulou12} we have argued that automatic
$\gamma$-ray quenching, i.e. 
a loop of processes which result in spontaneous $\gamma-$ray 
absorption accompanied by production of electron-positron pairs and soft
radiation, can become instrumental in the modelling 
of high energy sources. 
Its application to any $\gamma$-ray
emitting region, makes it relevant
to both leptonic and hadronic models; the fact, however, that quenching
requires rather high magnetic fields makes it more relevant
for the latter. In the present paper we revisit
the hadronic model taking into account the effects of
non-linear photon quenching.
This, as we shall show, has as a result to
exclude many sets of parameters, which otherwise
would give good fits to observations,
as the non-linear cascade growth modifies drastically the
produced multiwavelength spectrum. 

As a typical example we focus on the quasar 3C 279. 
This has been detected for the first time in very high-energy (VHE) $\gamma$-rays at $> 100$ GeV by the MAGIC telescope 
\citep{albert08} and
ever since it remains the most distant VHE $\gamma$-ray source with a well measured redshift. 
Both leptonic and (lepto)hadronic models have been applied \citep{boettcher09} and it has been pointed out that 
the former require the system to be well out of equipartition.  

In the present work 
we fit only the VHE part of the spectrum using a proton-synchrotron blazar model and use the contemporary X-ray data (see e.g.
\cite{chatterjee08}) as an upper limit;  
the modelling of the complete 
multiwavelength spectrum
requires a primary leptonic population radiating in the IR/X-ray energy bands
which, we assume, that can always be determined.
We show that one can derive a lower limit for the Doppler
 factor of the radiating blob by just combining (i)  the values of the proton luminosity and the Doppler factor for a given 
magnetic field that provide a good fit to the MAGIC data and (ii) the fact that for 
high values of the proton injection compactness the absorption of $\gamma$-rays becomes non-linear leading to an 
overproduction of softer photons
and consequently violating the X-ray observations {\sl even in the absence of a leptonic component}.
We also show that the minimum possible value of the Doppler factor is largely independent of
the magnetic field and this has some interesting implications for the energetics of 3C 279.

The paper is structured as follows. In \S2 we give some simple, qualitative estimates on the 
effects of quenching on fiting the multiwavelength data of blazars, in \S3 we apply these,
using a numerical code, to the case of 3C~279 and we conclude in \S4 with a discussion of 
our results.
For numerical results we adopt working in the $\Lambda$CDM cosmology with
$H_0=70$ km s$^{-1}$ Mpc$^{-1}$, $\Omega_{\rm m}=0.3$ and $\Omega_{\Lambda}=0.7$. 
The redshift of 3C 279 $z=0.536$ corresponds to a luminosity distance $D_{\rm L}=3.08$ Gpc.

\section{Analytical esimates based on automatic quenching} 

In this section we will justify the existence of a minimum Doppler factor
using analytical expressions in the simplest possible framework.
 The functional dependence of various physical quantities on the magnetic field strength, 
such as the particle density and the observed jet power, will also be derived. 

We consider a spherical blob of radius $R$ moving with a Doppler factor
$\delta$ with respect to us and containing a magnetic field of strength $B$. 
We further assume that ultra-relativistic protons with a power law distribution of index $s$
are being constantly injected into the source with a rate given by
\eqb
\tilde{Q}_{\rmn p}=\tilde{Q}_{\rmn {po}} \gp^{-\rmn s} H(\gp-\gmn) H (\gmx-\gp),
\label{injection}
\eqe
where $\gmn$ and $\gmx$ are the lower and upper limits of the injected distribution respectively
and $H(x)$ is the Heaviside function. 
 $\tilde{Q}_{\rmn {po}}$ is  the normalization constant and is also directly related 
to the proton injection compactness as:
\eqb
\lp & = & \tilde{Q}_{\rmn {po}} m_{\rmn p} c^2 \frac{\sth R }{3 \tcr} \frac{\gmn^{-\rmn s+2}-\gmx^{\rmn s+2}}{2-s}, \quad (s\ne 2) \cdot
\eqe
or
\eqb
\lp & = & \tilde{Q}_{\rmn {po}} m_{\rmn p} c^2 \frac{\sth R }{3 \tcr} \ln \left(\frac{\gmx}{\gmn} \right), \quad (s= 2),
\eqe
where $\tcr=R/c$. 
This can be further related to the proton injected luminosity by the relation
\eqb
L_{\rm p} & = & \frac{4\pi R \mpr c^3}{\sth}\lp,
\eqe
where $\sth$ is the Thomson cross section.

Protons will lose energy by synchrotron radiation,
photopair and photopion processes as has been described in 
\cite{petropoulou12}
 and therefore the proton distribution 
function is given by the solution of a time-dependent kinetic
equation that includes particle injection in the form of equation (\ref{injection})
in addition to particle losses and escape. 
Furthermore, since
the loss processes will create photons and electrons, 
one has to follow the evolution of these two species 
by writing two additional kinetic equations for them.
The solution of the system of the resulting three coupled 
partial integrodifferential equations gives the corresponding particle
distribution functions and the multiwavelength photon spectra can be calculated in a straightforward manner.
The picture above is rather complicated and it can be treated only numerically (see section 3).

In the case, however, where the cooling of protons does not play a significant role in comparison to particle escape, 
the steady-state proton number density is then simply given by
\eqb
n_{\rmn p}(\gamma) = \tcr \tilde{Q}_{\rmn {po}}\gamma^{-s}, \textrm {for} \ \ \ \gmn \le \gamma \le \gmx ,
\label{np}
\eqe
where we have assumed that the proton escape timescale equals the crossing timescale. 
Moreover, since we fit only the VHE $\gamma$-rays and not the whole multiwavelength spectrum, there is 
no need in calculating the evolution of the secondary leptons that will radiate at lower energies.
MAGIC data lie above $10^{25}$ Hz. The $\gamma$-ray spectrum in $\nu F_{\nu}$ peaks
at $\epsilon_{\gamma}^{\rm obs}=100$ GeV or equivalently at frequency $\nu_{\gamma}^{\rm obs}=2\times 10^{25}$ Hz\footnote{Throughout the present work 
quantities with the index `obs' will refer to the observer's frame, whereas all other
quantities refer to the blob frame.},
 which we will try to fit with 
 the proton synchrotron spectrum emitted by the particle distribution of equation (\ref{np}). 
We will focus on the case where the distribution of protons is not cooled
 and the synchrotron spectrum peaks at the maximun synchrotron energy, which implies that
$t_{\rm cool}(\gmx)\le \tcr$.
 Thus, our analysis is valid
under the prerequisite
\eqb
B^2 \gmx \lesssim 10^{-3} R \ \ \textrm{in cgs}.
\label{tcool}
\eqe
In this framework we obtain our first relation
\eqb
\eg^{\rm obs} = C_1 \delta B \gmx^2,
\label{eg}
\eqe
where $\eg^{\rm obs}=h \nu_{\gamma}^{\rm obs}$  and $C_1=\hbar e/ \mpr c (1+z)$. The total synchotron power per frequency in
the blob frame for the proton distribution of equation (\ref{np}) is given by  
\eqb
P_{\nu }=C_2 N_{\rm p} B^{\frac{s+1}{2}} \nu^{-\frac{s-1}{2}},
\label{pn}
\eqe
where 
\eqb
C_2 \approx \frac{2e^{7/2}}{\sqrt{6 \pi} \mpr^{3/2} c^{5/2}}
\eqe
and $N_{\rm p}$ is the total number of radiating protons in the blob, which
is given by $N_{\rm p}=\tcr  V \tilde{Q}_{\rmn {po}}(1/\gmn - 1/\gmx) \approx \tcr V \tilde{Q}_{\rm po}$,
where the approximation $\gmx^{-1} << 1$ has been made; here we have used also $\gmn=1$ and $s=2$.
This can also be expressed in terms of proton energy density $\up$ as
\eqb
N_{\rm p}= \frac{\up V} { \ln(\gmx) \mpr c^2}.
\eqe
In the above expressions $V\approx \pi R^3$ is the volume 
of the spherical blob.

\subsection{Derivation of the minimum Doppler factor}

By combining equation (\ref{eg}) with the integrated power over all synchrotron frequencies
(equation \ref{pn}) one obtains the total comoving synchrotron power 
\eqb
L=2 C_2 N_{\rm p} (\nu_{\gamma}^{\rm obs})^{1/2} B^{3/2} \delta^{-1/2}
\label{Lg-co}
\eqe
that is related to the total observed $\gamma$-ray luminosity $L_{\gamma}^{\rm obs} \approx 10^{48}$ ers/s by the
usual relation $L_{\gamma}^{\rm obs}=\delta^4 L$. This leads to our second relation
\eqb
L_{\gamma}^{\rm obs}= C_2' \up B^{3/2} \delta^{7/2},
\label{Lg}
\eqe
 where $C_2'=2 C_2 V (\nu_{\gamma}^{\rm obs})^{1/2}/ \ln(\gmx) \mpr c^2$.
For a specific magnetic field 
 we find a relation of proportionality
\eqb
\delta \propto \up^{-2/7} \propto \lp^{-2/7},
\label{d-lp}
\eqe
which is also verified by the detailed numerical treatment (see section 3.2).
We note that the proportionality at the right hand side of relation (\ref{d-lp})
holds only if the magnetic field is not strong enough to 
 cause significant proton cooling
due to synchrotron radiation.

The maximum proton Lorentz factor has remained so far undetermined. 
However, the fact that the proton gyroradius $r_{\rm g}$ should be less or equal than
the size of the blob $R$ \citep{hillas84} sets a strong upper limit

\eqb
\gmx=\kappa \frac{e B R}{ \mpr c^2},
\label{rg}
\eqe
where 
$\kappa$ is a scaling factor that takes values less or equal to unity. It
has been introduced in order to take into account the fact, that good fits can be obtained for certain parameter
sets with $\gmx < < e B R / \mpr c^2$; as an example see the parameters
used for figure \ref{noquench}. 
By combining the above relation with equation (\ref{eg}) we find a lower limit for the Doppler
factor
\eqb
\delta \ge \dmin= C_3 B^{-3},
\label{dmin1}
\eqe
where 
\eqb
C_3=\frac{1}{C_1}\left( \frac{ \mpr c^2}{ \kappa e} \right)^2 \frac{\eg^{\rm obs}}{R^2}.
\eqe

Automatic quenching of $\gamma$-rays was 
irrelevant to the derivation of the relations presented up to this point. Actually, it will
not play any role in the evolution of the system if the magnetic field is weak enough. There is,
in other words, a necessary but not sufficient condition for the operation of automatic quenching, the 
so-called \textit{feedback criterion}. This can be derived from the requirement that the magnetic field is strong
enough so that the synchotron photons of the automatically produced pairs lie above the threshold for further photon-photon
absorption on the $\gamma$-rays \citep{SK07, PM11, petropoulou12}. This requirement
sets a lower limit for the magnetic field 
\eqb
B\ge 8\Bcr\left(\frac{\me c^2}{\epsilon_{\gamma}}\right)^3,
\label{feedback}
\eqe
where $\Bcr=4.4\times 10^{13}$ G. 
If we set $\epsilon_{\gamma}=\eg^{\rm obs}(z+1) \delta^{-1}$ and use equations (\ref{eg}) and (\ref{rg})
we find an expression for the magnetic field $\Bq$, below which the feedback criterion is not satisfied.
This involves only physical constants except for the size of the blob $R$ and the scaling factor $\kappa$:
\eqb
\Bq=(8\Bcr)^{1/10} \left(\frac{ \mpr \me c^3}{ \hbar e}\right)^{3/10} \left(\frac{ \mpr c^2}{e\kappa R} \right)^{3/5}.
\label{Bq}
\eqe
In order to have an estimate,  the above expression  gives $\Bq \approx 3.5$ G
for $R=3\times 10^{16}$ cm and $\kappa=1$.
Note that for a given $R$ this is also the minimum value of $\Bq$. 
One could also express $\Bq$  in terms of $\gmx$ as
\eqb
\Bq = (8 \Bcr)^{1/4} \left(\frac{\mpr \me c^3}{ \hbar e}\right)^{3/4} \gmx ^{-3/2},
\label{Bqgmx}
\eqe
as long as $\gmx$ does not exceed the value given by equation (\ref{rg}) for $\kappa=1$.
From equation (\ref{Lg}) becomes evident 
that if the Doppler factor decreases then the proton energy density or the proton
compactness equivalently should increase in order to obtain the same observed luminosity. 
If, however, the feedback criterion is satisfied, then the proton compactness is bounded from above.
This immediately sets a lower limit for the Doppler factor. We proceed next to obtain an expression of $\dmin$.
In previous work we have obtained an analytical
 expression for the critical $\gamma$-ray compactness (see equation (34) of \cite{PM11}), which will prove very useful for our analytical
calculations
\eqb
\lcr \approx C_4 B^{1/2} \delta^{-1/2},
\label{lcr}
\eqe
where $C_4= \left( 2\eg^{\rm obs} / \sigma_0^2 \Bcr \me c^2\right)^{1/2}$ and $\sigma_0=4/3$ is a normalization constant.
The $\gamma$-ray compactness is defined as $\lgg= L \sth / 4 \pi R \me c^3$.
Using equation (\ref{Lg-co}) expressed in terms of proton compactness $\lp$ instead
of total proton number $N_{\rm p}$ we can rewrite the above definition as
\eqb
\lgg=C_5 \lp B^{3/2} \delta^{-1/2},
\label{lg}
\eqe
where 
\eqb
C_5= \frac{6C_2 V (\nu_{\gamma}^{\rm obs})^{1/2}}{4 \pi R^2 \me c^3 \ln(\gmx)}.
\eqe
In order to avoid an overproduction of soft photons that would violate the upper limit
set by the X-ray observations we require that $\lgg\le \lambda \lcr$, where $\lambda$ is a numerical
factor between 1 and 10; the exact value can be estimated only 
numerically. This requirement
defines a maximum proton compactness
\eqb
\ell_{\rm p, max}= \lambda \frac{C_4}{C_5}B^{-1},
\label{lpmax}
\eqe
which, if inserted in equation (\ref{Lg}), gives the minimum Doppler factor
\eqb
\dmin=\left( \frac{L_{\gamma}^{\rm obs} \sth }{4 \pi R \lambda \me c^3 C_4} \right)^{2/7} B^{-1/7} .
\label{dmin2}
\eqe
We note that $\dmin$ given by the equation above is a rather
robust limit, since it depends weakly on the physical quantities.
We note also that the numerical factor $\lambda$ appears in the above 
expression raised to the power $-2/7$ and therefore
it does not affect the value of $\dmin$ severely. 

Summarizing, we have found that for $B< \Bq$, the minimum value of
the Doppler factor is defined
by the requirement that the gyroradius is less than the blob size and it has a very strong dependence on
the magnetic field, i.e. $\dmin \propto B^{-3}$. On the other hand, for $B \ge \Bq$ , it
is defined by the requirement that the proton compactness cannot exceed
a critical value. 
The dependence on the magnetic field is in this case weak, i.e. $\dmin \propto B^{-1/7}$.

\subsection{Conditions for equipartition}

Estimations of the equipartition magnetic field of a relativistically moving blob are
common in the literature and are based on fitting the low-frequency synchrotron spectrum
of a power-law distribution electrons to radio observations \citep{pacholczyk70}.
Here we derive an expression for the 
equipartition magnetic field in a proton-synchrotron
blazar model, replacing the power-law distribution 
of electrons with protons and the radio with VHE $\gamma$-ray
observations.
From equation (\ref{Lg}) we find that the comoving proton energy density is
given by
\eqb
\up = A B^{-3/2} \delta^{-7/2},
\label{up}
\eqe
where $A= L_{\gamma}^{\rm obs} {C_2'}^{-1}$.
Assuming that the energy density of secondary leptons is negligible with respect
to  that of protons, we can compute
the equipartition field $\Beq$
\eqb
\Beq  =  (8 \pi A)^{2/7} \delta^{-1}
\label{beq0}
\eqe
or 
\eqb
    \Beq = 718 \ L_{\gamma,48}^{\rm obs} R^{-6/7}_{16} (\nu_{\gamma,25}^{\rm obs})^{-1/7} 
\ln(\gamma_{\rmn max, 10}) \delta^{-1} \quad \textrm {in G} .
\label{beq}
\eqe
We note that here and in what follows the convention $Q_X \equiv Q/10^X$ in cgs units was adopted unless stated otherwise.
The above expression apart from the numerical constant, is identical to that found 
using  equipartition arguments in the context of a leptonic model (see e.g. equation (A7) of \cite{harris02}).
In the previous subsection it was found that the Doppler factor of the blob
has a minimum value $\dmin$ and its functional dependence on $B$ was also
derived. One can therefore investigate whether or not
the system can achieve equipartition 
under the requirement that $\delta=\dmin$
and in this case estimate $\Beq$ using equation (\ref{beq}).

\begin{figure}[h]
 \centering
\resizebox{\hsize}{!}{\includegraphics{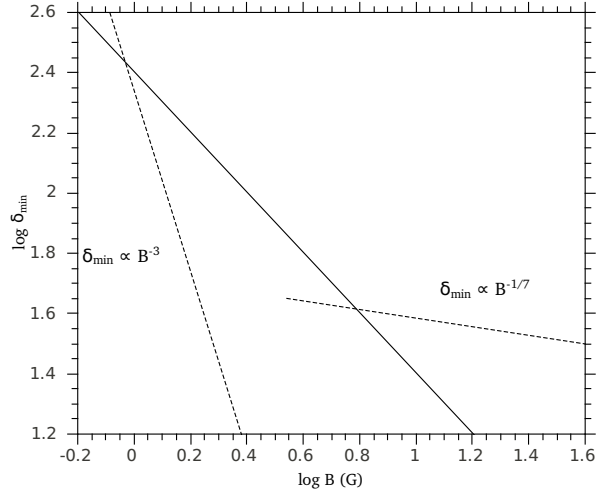}}
\caption{Minimum Doppler factor as a function of $B$ (dashed lines) and locus of points
satisfying the equipartition condition (solid line) for $L_{\gamma}^{\rm obs}=10^{48}$ erg/s, 
$\nu_{\gamma}^{\rm obs}=2\times10^{25}$ Hz, $R=3\times10^{16}$ cm and $\lambda=5$.}
\label{dmin-ana}
\end{figure}

Figure ~\ref{dmin-ana} shows $\dmin(B)$ given by equations (\ref{dmin1}) and (\ref{dmin2}) and
the line $\delta \propto \Beq$ for $R=3\times10^{16}$ cm and $\lambda=5$. The discontinuity
occurs at $\Bq$, where the functional dependence of $\dmin$ on $B$ changes. 
The expressions derived
should not be trusted in the neighbourhood of $\Bq$. The points of intersection 
correspond to the equipartition magnetic field values. 
The physical system can be found in equipartition either moving
with a high Doppler factor and being weakly magnetized or moving with a modest Doppler factor and containing a 
stronger magnetic field. 

A robust result of our treatment, which also differentiates
it from other works, is that for a given magnetic field there is
a minimum value for the Doppler factor of the flow that is set either by
the gyroradius or by automatic quenching arguments.
Whenever the analysis of a physical problem leads in the
derivation of an extremum for some parameter, it is interesting to
study the properties of the physical system in this case. For this reason,
in the following 
we present how the particle energy density and the power of the jet depend on
the magnetic field in the particular case of $\delta=\dmin$. 
%
Using equations (\ref{dmin1}), (\ref{dmin2}) and (\ref{up}) 
we find that
\eqb
\up(\dmin(B)) \equiv \up^{\star} \propto  B^9 \quad \textrm{for} \quad B<\Bq
\label{up1}
\eqe
and
\eqb
\up(\dmin(B)) \equiv \up^{\star} \propto B^{-1} \quad \textrm{for} \quad B >\Bq. 
\label{up2}
\eqe
Note the large change in the dependence of the proton energy density on $B$.
The last proportionality is not valid for very high values of the magnetic
field, since in this case the synchrotron cooling becomes important and is against
our initial assumptions.

\subsection{Jet power estimates}

In an one-zone homogeneous model the emitting plasma is confined
in a blob with radius $R$ moving with a velocity $\beta c$ and bulk Lorentz
factor $\Gamma$ at an angle $\theta$ with respect to the line of sight. 
The radiation is further assumed to originate from a 
region in the jet of volume $V \approx \pi R^3$.
The energy densities of particles, magnetic field and radiation are contributing to the jet power.
Using the approximations $\theta \approx 1/\Gamma$, $\delta \approx \Gamma$ and $\beta \approx 1$ 
the observed 
jet power is given by
\eqb
\Pjet^{\rm obs}=\pi R^2 \delta^2 \beta c (\ub+u_{\rm part}+u_{\rm rad}),
\label{pjet-def}
\eqe
where $u_{\rm part} \approx \up$, $\beta =\sqrt{1-1/\Gamma^2}$
and  $u_{\rm rad}$ is the radiation energy density, 
which for simplifying reasons will not
be included in the following analytical calculations. It will be however 
taken into account in our numerical treatment of section 3. Note also that
in our analytical formulation we have not included any population of `cold' protons
and electrons, which  are physically implied by charge conservation arguments. However,
their contributions to the total jet power are minor (see for example \citealt{protheroe01})
and would not affect our results significantly. 

Inserting the expression for $\up$ given by equation (\ref{up})
in the above relation we find that the jet power is a function only of 
the product $\delta \cdot B$

\eqb
\Pjet^{\rm obs}=\pi R^2 c \left(\frac{(\delta B)^2}{8 \pi} + A (\delta B)^{-3/2} \right).
\label{pjet}
\eqe
The jet power is minimized when the following condition holds
\eqb
B = (6 \pi A)^{2/7} \delta^{-1}.
\label{pjet-min}
\eqe
Comparison of equations (\ref{beq0}) and (\ref{pjet-min})
shows that whenever the system achieves equipartition the jet power
also minimizes at the value
\eqb
P_{\rm jet, min}^{\rm obs} =5.8\! \times \! 10^{47}
(L_{\gamma,48}^{\rm obs} \ln(\gamma_{\rm max,10}) )^{\frac{4}{7}} \! \left(\frac{R_{16}}{\nu_{\gamma,25}^{\rm obs}}\right)^{\frac{2}{7}}
\! \! \! \textrm{erg/s}.
\label{pjetminval}
\eqe
Thus, even the minimum jet power obtained in the framework of the proton-synchrotron
blazar model is high compared to values inferred from pure leptonic modelling of blazar emission; see 
e.g. \cite{celotti08}. Since for a large range of $B$ values
the system lies far from equipartition (see figure~\ref{dmin-ana}), equations (\ref{beq}) and
 (\ref{pjet-min}) imply that the calculated jet power differs significantly from its minimum value.

The fact that the Doppler factor has a lower limit implies that for magnetic fields
higher than a certain value $B_{\rm h}$, equation (\ref{pjet-min}) cannot be further satisfied and there is
no set of parameters that lead to minimization of the jet power.
The characteristic value $B_{\rm h}$ can be derived by combining equations (\ref{dmin2}) and (\ref{pjet-min}). It is
given by:
\eqb
B_{\rm h} = 11.6 \ R_{16}^{-2/3} \lambda^{1/3} \ln(\gamma_{\rm max, 10})^{1/3} \ \textrm{ G}.
\label{bhigh}
\eqe
The same holds also for weak enough magnetic fields. 
Thus, by combination of equations (\ref{dmin1}) and (\ref{pjet-min}) one derives
a second characteristic value $B_{\rm l}$ for the magnetic field, below which there is no set of
parameters that could minimize the power of the jet
\eqb
B_{\rm l} = 0.96 \ R_{16}^{-4/7} (\nu_{\gamma,25}^{\rm obs})^{8/14} 
\left(L_{\gamma,48}^{\rm obs} \ln(\gamma_{\rm max, 10}) \right)^{-1/7} \ \textrm{G}.
\label{blow}
\eqe
Therefore, the discussion above reveals another important physical aspect of $\dmin$;
it limits the range of $B$-values among which one should choose
in order to obtain the minimum power of the jet.
If $\delta=\dmin$ the dependence of $\Pjet^{\rm obs}$ on $B$ can also be derived by combining equations
(\ref{dmin1}), (\ref{dmin2}) and (\ref{pjet})
\eqb
\Pjet^{\rm obs}(B;\dmin) \propto a_1 B^{-4} + a_2 B^{3} \quad \textrm{for} \quad B <\Bq
\label{pjet1}
\eqe
and
\eqb
\Pjet^{\rm obs}(B;\dmin)\propto b_1 B^{12/7} + b_2 B^{-9/7} \quad \textrm{for} \quad B >\Bq,
\label{pjet2}
\eqe
where $a_{1,2}$ and $b_{1,2}$ are constants. Note that the functional form of $\Pjet^{\rm obs}$
 is not trivial and it consists of four power-law segments. 
We will return to this
point in the next section, where we perform the numerical approach in fitting the $\gamma$-
ray observations of 3C 279. 

\section{Numerical results}

In what follows we will present a 
method of fitting the February 26, 2006 
 high energy observations of quasar 3C 279 using an one-zone hadronic model. 
Our working framework is analogous to that adopted
in the previous section but with
 two main differences: 
\begin{enumerate}
 \item[(i)]  no assumptions about the relative importance of
synchrotron cooling with respect to that due to
photohadronic processes for the proton distribution 
are made; particle
and photon distributions are self-consistently obtained as the solution
to a system of three coupled integrodifferential equations. This is
done with the help of the numerical code described in 
\cite{mastkirk95} and \cite{mastkirk05}.
\item [(ii)]  $\chi^2$ statistics was used to determine which
proton-synchrotron spectra provide `good' fits to the VHE $\gamma$-ray data.
\end{enumerate}
In order to minimize the number of free parameters
we keep fixed the following:
$\gmn=1$, $s=2$ and $R=3\times 10^{16}$ cm.
We note that in the analytical formulation of section 2, we 
have used explicitly the parameter $R$, although it combines an
 observable quantity, i.e. the variability timescale $t_{\rm var}$ with the Doppler factor
$\delta$. For this, the derived values of $\delta$ throughout the present work should be 
checked \textit{a posteriori} against 
the relation imposed by the variability of the source
\eqb
\delta \ge \frac{4 R_{16} (1+z)}{t_{\rm days}},
\label{tvar}
\eqe
where $t_{\rm days}$ is the observed variability timescale normalized to
1 day. One could instead work with the observable $t_{\rm var}$ and incorporate this extra constraint in his/her
analytical treatment. Possible effects of different adopted values for $R$ and $s$ on our 
results will be discussed in section  4.

\subsection{The method}
Our aim is to produce a parameter space which gives `good' fits to 
February 2006 observations of 3C 279. Application
of existing theoretical models to AGN observations
leads to one set of parameters that minimizes the 
$\chi^2$-value. However, the goodness-of-fit may pose some interesting questions,
especially when two very different 
sets of parameters may have very similar
values of $\chi^2$. 
For this reason, in what follows,
 we do not restrict ourselves
to the best fit with the minimum $\chi^2$-value, but we rather relax
the definition of a `good' fit.  Thus, fits to the TeV data
that have $\chi^2_{\rm red}<1.5$ are characterized as `good' and they
are obtained for different combinations of the parameters. This, as we will show in
the next section, results in the formation of a parameter space instead of a single 
set of accepted parameter values.

Since we keep $\gmn$, $s$ and $R$ fixed the number
of free parameters in the context of a pure hadronic model reduces to 
four: $B$, $\gmx$, $\lp$ and $\delta$. Thus, we search for combinations of 
the aforementioned parameters that provide good fits to the TeV data.
At the same time we have treated the X-ray observations as upper limits:
As long as the proton induced emission is below the X-ray data,
we assume that one can always find a fit to them by using a
suitably parametrized leptonic component. 
On the other hand, if the  emission due to photon quenching is above the X-rays,
then we discard the fit. The steps of the algorithm followed are

\begin{enumerate}
\item Choose a value for the magnetic field strength $B$.
\item  Choose a value for the maximum proton Lorentz factor $\gmx$ starting
from the highest possible value, which is
 imposed by the Hillas criterion (see equation (\ref{rg})).
\item  Choose an injection compactness for the proton
distibution $\lp$. For high enough values, automatic quenching
sets in and creates a soft-photon component that exceeds the X-ray observations.
\item  Choose a value for the Doppler factor $\delta$ after taking into account that
the variability timescale of 3C 279 does not exceed that of one day.
\end{enumerate}

\begin{figure}
 \centering
\resizebox{\hsize}{!}{\includegraphics{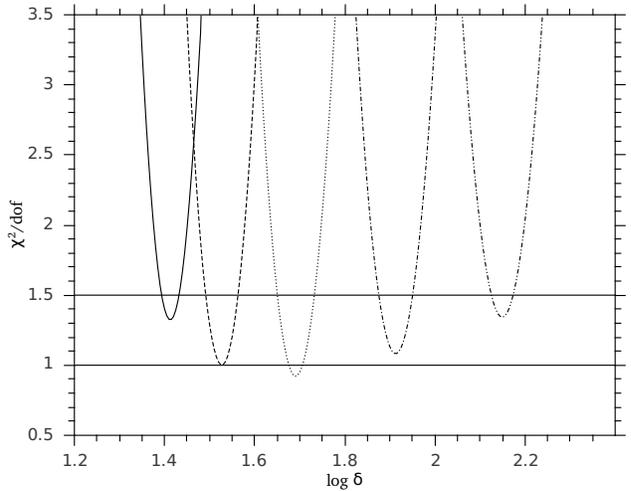}}
\caption{Reduced $\chi^2$ as a function of the Doppler factor $\delta$ for $B=40$ G,
$\gmx=4\times10^9$ and $\lp=10^{-8}$ (solid line), $\lp=10^{-7}$ (dashed line),
 $\lp=10^{-6}$ (dotted line), $\lp=6.3\times10^{-6}$ (dashed-dotted line) and 
$\lp=2.5\times10^{-5}$ (dashed-double dotted line).
The horizontal solid line with $\chi^2_{\rm red}$ shows the upper 
limit below which a fit is characterized as  `good'. The line $\chi^2_{\rm red}=1$
is also shown.
}
\label{chi2}
\end{figure}

\begin{figure}
\centering
\includegraphics[width=10cm, height=9cm]{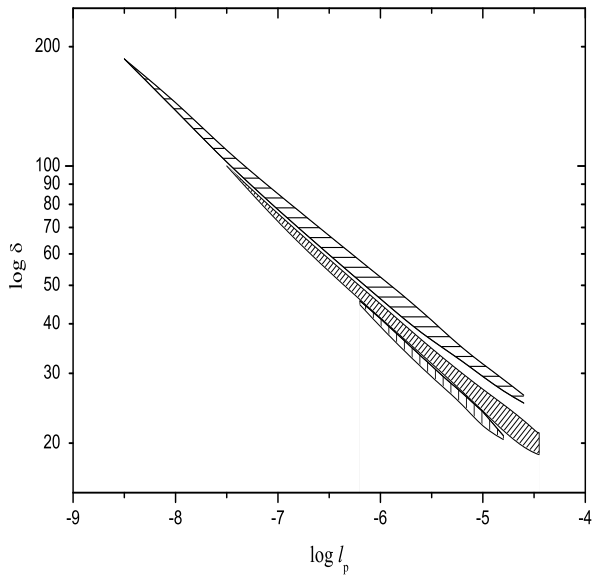}
\caption{Parameter space of pairs ($\lp$,$\delta$) for $B=40$ G and different $\gmx$ that
correspond to TeV fits with reduced $\chi^2<1.5$. 
Each shaded area corresponds to a different value of $\gmx$ that ranges from $10^{9.6}$ to $10^{10}$
with a step of 0.2 in logarithmic units.
For each value of $\lp$
two values of $\delta$ can be 
determined, which form the envelope
of each striped area. The shaded areas enclosed by the solid lines
correspond also to values of $(\lp,\delta)$ for which food fits to 
the TeV data ($\chi^2_{\rm red}<1.5$) can be obtained.}
\label{space}
\end{figure}

For each triad $(B,\gmx,\lp)$ two values of $\delta$ can be found that correspond
to fits with $\chi^2_{\rm red}\le 1.5$. This can be easily explained
by the parabolic shape of the $\chi^2_{\rm red}$ curves.
Figure \ref{chi2} shows 
$\chi^2_{\rm red}$ as a function of $\delta$ for $B=40$ G and $\gmx=4\times 10^9$.
Different curves correspond to different values of $\lp$. The part of the curve that lies
below the horizontal line with $\chi^2_{\rm red}=1.5$  provides  good fits. Its projection on the horizontal axis
defines an interval of $\delta$-values. For 
 simplifying reasons, we consider that each monotonic branch of the $\chi^2_{\rm red}$-curve that lies
below the horizontal line is
represented only by one $\delta$-value, i.e. only by one point
on the horizontal axis.   Thus, for each $\lp$ we find two 
representative values of the Doppler factor.
The error we make in this case is not large since
the curves are very steep.  Note also that Doppler factors lying between the two
representative $\delta$-values also provide good fits.
If we repeat the above proceedure for various $\gmx$ we can then
create a parameter space for $\lp$ and $\delta$ for a specific value of the $B$-field. 
An example is shown in figure \ref{space} for $B=40$ G and different $\gmx$.
For the case considered here,  $\gmx$ ranges from
$10^{9.6}$ (horizontally dashed area) to  $10^{10}$ (vertically
dashed area) with a step of 0.2 in logarithmic units.
The envelope of each striped area is the result of the 
two representative values of $\delta$ defined for each $\lp$, as previously described.
Note that in a log-log plot the lines of the envelope are power laws with 
an exponent $-0.22$, i.e. $\delta \propto \lp^{-0.22}$.
 This almost coincides with the value $-2/7$ found
in our analytical approach (see equation \ref{d-lp}). 
Values of $\gmx$ outside this range 
do not provide good fits and are therefore discarded. Thus, the shaded areas in 
Figure \ref{space} depict all the possible combinations of $\lp$, 
$\gmx$ and $\delta$ that provide good fits to the TeV observations.
We note that one can use low values of the proton injection compactness
$\lp$ and get acceptable TeV fits but this can be done only if he/she
allows the Doppler factor to take high values
-- this is due to the fact that the observed and blob frame luminosities are related by
$L_{\rm obs} \propto \delta^4 L_{\rm int}$, while $L_{\rm int}\propto L_{\rm p} \propto \lp$.
On the other hand,
higher values of $\lp$ naturally result in lower $\delta$ values. 
However, as stated earlier,
 we find that 
$\gamma-$ray quenching plays an important role when $\lp$ takes higher values.
In this case the spontaneously
produced soft photons increase and eventually will start violating  
the X-ray observations, destroying the goodness of the fit. 
Therefore, an upper limit is imposed on the allowed values of $\lp$
which implies in turn,  the existence of a minimum value of the
Doppler factor (see equations (\ref{lpmax}) and (\ref{dmin2})).

\subsection{Minimum Doppler factor}

\begin{figure}
 \centering
\resizebox{\hsize}{!}{\includegraphics{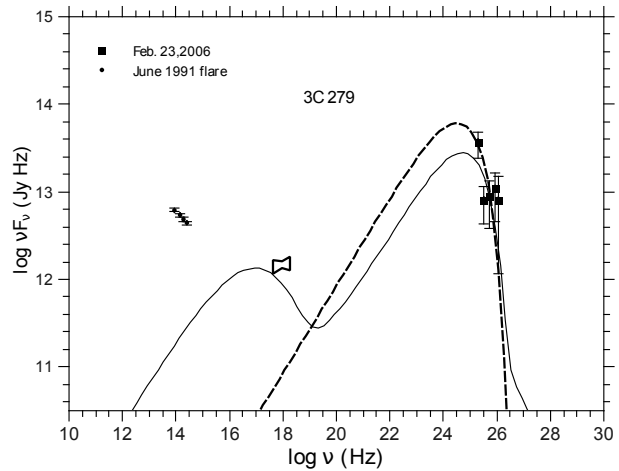}}
\caption{Multiwavelength spectra of 3C 279 in the context of a pure hadronic model for 
$R=3\times 10^{16}$ cm, $B=40$ G, $\gmn=1$, $\gmx=6.3\times10^9$ and $s=2.0$. The solid curve is
obtained with all processes taken into account for $\lp=10^{-4.4}$ and $\dmin=18.6$, whereas the dotted one
is obtained for $\lp=10^{-3}$ and $\delta=10$ after artificially ignoring $\gamma \gamma$ absorption.
Squares represent the VHE-detection by MAGIC \citep{albert08} after correcting for intergalactic $\gamma \gamma$ absoprtion,
 the bowtie represents the Swift data and filled circles are IR data during the flare of June 1991 
taken by \citep{hartman96}.}
\label{noquench}
\end{figure}

Figure \ref{noquench} exemplifies the effect of $\gamma$-ray absorption 
on the adopted values of $\lp$ and $\delta$ for fitting the TeV data.
The MAGIC TeV data points show the measured flux corrected for intergalactic
$\gamma \gamma$ absorption. The lowest possible level of
extragalactic background light (EBL) according to \cite{primack05} has been used
for the de-absorption of the VHE $\gamma$-rays.
The full line curve depicts the last good-fit spectrum we obtain 
for a high $\lp$. Here $\lp=4\times10^{-5}$, $\gmx=10^{9.8}$
and $\delta_{\rm min}=18.6$.  Other parameters used for this plot are: $\gmn=1$, $R=3\times10^{16}$ cm,
$B=40$ G and $s=2$. Note first that the minimum Doppler factor derived here does not violate the
inequality (\ref{tvar}) for $t_{\rm days}=1$, and second that $B^2 \gmx$ satisfies the constraint given by relation (\ref{tcool}). 
The non-linear cascade due to quenching
has produced a soft component that is almost at the observed
X-ray flux. 
Any attempt to increase $\lp$ further will produce a higher flux of soft
photons which would violate the X-rays.
To demonstrate this point further,
we have ran the code for 
a high value of $\lp=10^{-3}$. This value is well inside the quenching regime
and thus should be excluded due to strong X-ray production. However, by turning
artificially $\gamma\gamma$ absorption off, we inhibit the growth of the
non-linear cascade and therefore a good fit can be obtained with a $\delta$ as low as 10 (dashed line).
Moreover, if $\gamma \gamma$ absorption is
treated in an approximate semi-analytic manner as a linear absorption process,
automatic quenching of $\gamma$-rays  will not even occur , since it is a purely non-linear absorption
process. Therefore, no limiting value of the proton and $\gamma$-ray compactness would be found and
the transition  of the hadronic
system to supercriticality would not be seen. As a result one would find
erroneously a more extended parameter space that fits the data. 
In such case, the Doppler factor would not be limited to a lower value apart from the one
found by the usual gyroradius arguments made in section 2.

\begin{figure}
 \centering
\resizebox{\hsize}{!}{\includegraphics{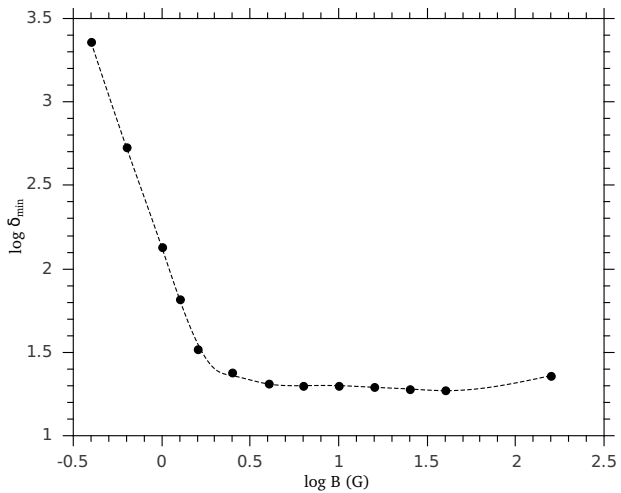}}
\caption{Minimum value of the Doppler factor $\dmin$ (points) as a function of the magnetic field strength $B$. The dashed line
is the result of spline interpolation between the points.}
\label{dmin-num}
\end{figure}

As a next step we have obtained $\delta_{\rm min}$ for various values of the magnetic
field $B$. This is depicted in figure \ref{dmin-num}, where the two branches derived analytically
in the previous section (see figure \ref{dmin-ana}) are clearly seen. The power law dependence
of $\dmin$ on $B$ can be modelled as
of $\dmin \propto B^{-\alpha}$ with $\alpha \simeq 2.85$ for the \textit{low-B} branch and $\alpha
\simeq 0.05$ for the \textit{high-B} branch. We note that the numerically derived power-law exponents of
the two branches are very close to those given by equations (\ref{dmin1}) and (\ref{dmin2}) respectively.
A new feature of figure \ref{dmin-num} is that for very strong magnetic fields, where synchrotron cooling
affects a significant part of the proton power-law distribution, the minimum Doppler factor required for
a good fit slightly increases with increasing $B$.

\subsection{Energetics}
The previous analysis makes evident that acceptable fits 
to the TeV observations of 3C 279, 
within the context of the hadronic model,  
can be found for a large choice of magnetic fields,
maximum proton energies and injection luminosities. We
note also that these values are similar to the ones found 
in standard hadronic modelling of TeV sources \citep{boettcher09}.
 
\begin{figure}
 \centering
\resizebox{\hsize}{!}{\includegraphics{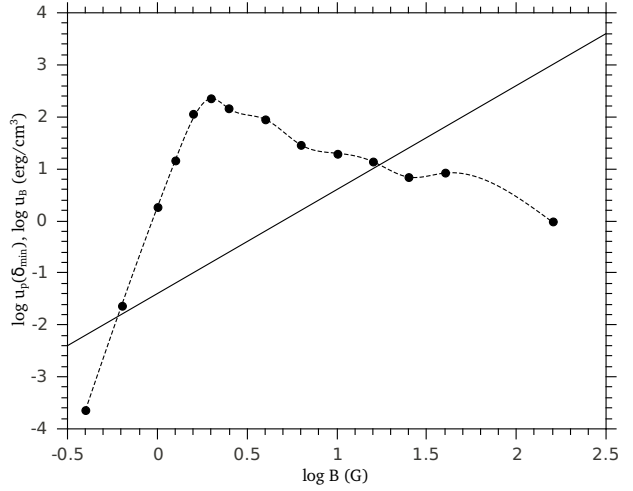}}
\caption{Logarithmic plot of the proton energy density in the case of the 
mimimum Doppler factor $\up^{\star}$ (points) 
and of the magnetic energy density (solid line) as a function of the magnetic field. 
Spline interpolation between the points results in the dashed line. 
}
\label{densities}
\end{figure}

Having already derived $\dmin$ for each magnetic field strength, we numerically then derive
the steady state proton energy density $\up^{\star}\equiv \up(\dmin(B))$ and plot 
this quantity as a function of $B$ -- see figure \ref{densities}.
The system becomes magnetically dominated either for high enough ($B > 10$ G) or low enough
 ($B < 0.7$ G) magnetic fields. These values
should be compared to $B_{\rm h}$ and $B_{\rm l}$ given by equations (\ref{bhigh}) and (\ref{blow}) respectively.
The two curves intersect at two points, exactly as was
analytically derived (see section 2.2). At these points  
the two energy densities are equal and this corresponds to a minimum
of the total energy density, i.e. to an equipartition between the particles and the magnetic field -- 
 we emphasize
that the quantity $\up^{\star}$ is not the minimum particle energy density.
 The fact that $\dmin$
has a strong dependence on $B$ at low values is also reflected to $\up^{\star}$.  
Our analytical result given by equation (\ref{up1}) should be compared to the numerical one, i.e. $\up^{\star} \propto B^{8.7}$.
For $B>2$ G (c.f. $B_{\rm q}$ of equation \ref{beq}) the particle energy density decreases with increasing magnetic field as $B^{-1.4}$. This
trend was also found analytically. The power-law exponent however is slightly different -- see equation (\ref{up2}).

\begin{figure}
 \centering
\resizebox{\hsize}{!}{\includegraphics{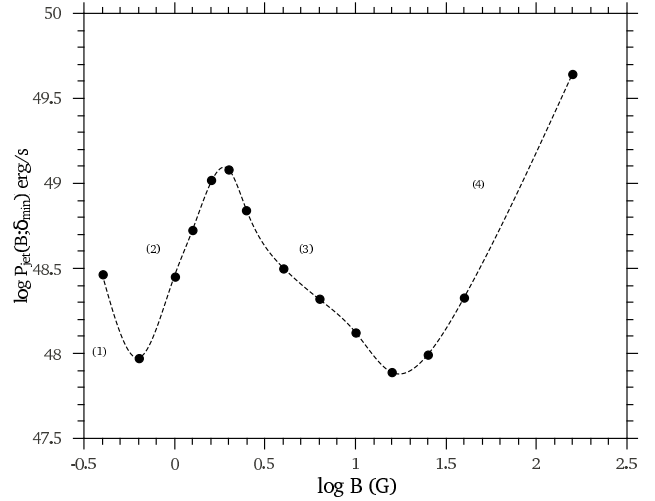}}
\caption{Jet power as a function of the magnetic field for $\delta=\dmin$ (points). The dashed line is
the result of spline interpolation. Numbers label the different power-law segments of the curve.
}
\label{jetpower}
\end{figure}

From equation (\ref{pjet-def}) and for $\delta=\dmin(B)$ and $\lp=\lp(\dmin)$ we have calculated numerically
the observed jet power, which becomes a function only of the magnetic field, i.e. $\Pjet^{\rm obs}(\dmin(B))$. 
This is shown in figure \ref{jetpower}. The function $\Pjet^{\rm obs}(\dmin(B))$ shows two local minima 
for two values of the magnetic
field that differ more than one order of magnitude. 
For a wide range of $B$ values the calculated jet power is 
rather high $\approx 10^{48}-10^{49}$ erg/s in comparison to leptonic models \citep{celotti08}. 
In a log-log plot the curve consists of four distinct power-law segments labeled with the numbers
1 to 4. In Table 1 the power-law exponents of the four segments and the corresponding analytically derived
values given by equations (\ref{pjet1}) and (\ref{pjet2})  are listed. Apart from the first segment both results are in good agreement.

\begin{table}
 \centering
\begin{tabular}{ c c c}
 \hline
\phantom{a} & \multicolumn{2}{c}{exponent}\\
\hline 
Number of segment & numerical & analytical \\
\hline \hline
(1)  & -2.50     &  -4.00  \\
(2) &  +2.30   &  +3.00 \\
(3) &  -1.30   &  -1.28\\
(4) &  +1.70   &  +1.71\\
\hline
\end{tabular}
\caption{Power-law exponent of $\Pjet^{\rm obs}(\dmin(B))$ as derived from the numerical and analytical treatment.}
\end{table}

\begin{figure}
 \centering
\resizebox{\hsize}{!}{\includegraphics{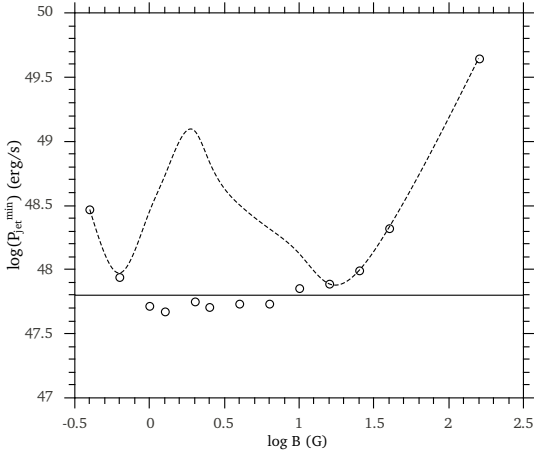}}
\caption{Minimum jet power value (open circles) as a function of the magnetic field strength. 
The
 horizontal line denotes the analytically derived minimum value of the jet power (equation \ref{pjetminval}) and the 
dashed line, which is same as in figure \ref{jetpower}, is shown for comparison reasons.
Open circles that lie on the dashed line correspond to the jet power calculated
for $\dmin$, since $\delta<\dmin$ would be required to minimize the jet power. }
\label{minpjet}
\end{figure}

\begin{figure}
 \centering
\resizebox{\hsize}{!}{\includegraphics{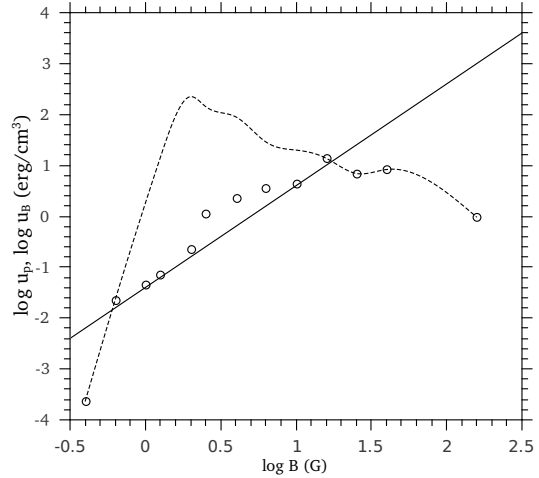}}
\caption{ Proton energy density (open circles) as a function of the magnetic field strength,
for those parameter sets that correspond to the minimum jet power. The magnetic field energy density
is shown with the solid line. The remaining symbols are the same as in figure \ref{minpjet}.}
\label{upub2}
\end{figure}

We have already shown that the jet power minimizes whenever $\delta$ and $B$ satisfy equation (\ref{pjet-min}). In this case
the system is close to a 
state of equipartition. We proceed to investigate whether these results are verified numerically.
For each value of $B$, we therefore search 
among the triads $(\gmx,\lp,\delta)$ found during
the fitting process for that specific set of values that minimizes the jet power.
Figures \ref{minpjet} and \ref{upub2} show respectively the minimum calculated jet power $P_{\rm jet, min}^{\rm obs}$ and 
the corresponding proton energy density as a function of $B$. 
Dashed lines are same as in figures \ref{densities} and 
\ref{jetpower}  and are shown for comparison reasons.

Figure \ref{minpjet} shows that for $ 0.7 \le B \le 10$ G all points  lie close to the minimum value 
$P_{\rm jet, min}^{\rm obs} \simeq 6.5\times10^{47}$ erg/s, whereas for $B > 10$ G and 
$B < 0.7$ G the minimum jet power found from our
numerical data sets coincides with that calculated for $\delta=\dmin$ (open circles lying on the dashed line). 
As already discussed in section 2.3 the jet power
calculated in this case is not the lowest possible minimum value, since this would be obtained for Doppler factors less than $\dmin$.
Note that the transitions occur at $B  \approx B_{\rm l} \approx 0.7 \ \textrm{G}$ and $B  \approx B_{\rm h} \approx 10 \ \textrm{G}$.

In figure \ref{upub2} one can see that $\up \approx \ub$ for $B_{\rm l} \le B \le B_{\rm h}$, 
which is in complete agreement 
with the analytical estimates of
section 2.3 . The fact that the calculated jet power is not the minimum possible one, can also be verified
by the fact that the system for $B > 10$ G and $B<0.7$ G is far from an equipartition state.

\begin{figure}
 \centering
\resizebox{\hsize}{!}{\includegraphics{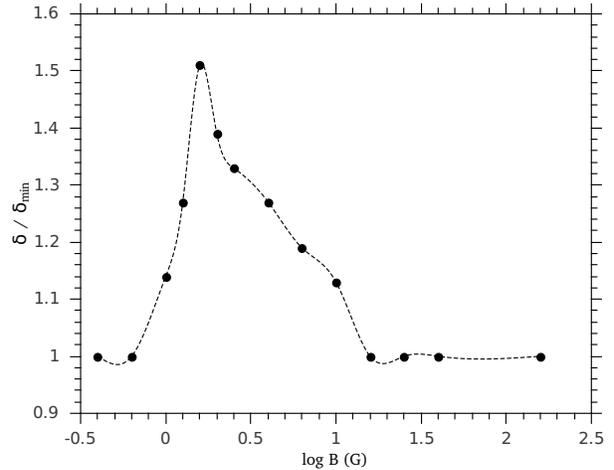}}
\caption{Ratio $\delta / \dmin$ as a function of the magnetic field.
Here $\delta$ is the Doppler factor required for minimization of the jet power. Interpolation between the points
results in the dashed line. }
\label{doppler}
\end{figure}

Each of the points shown as open circles in figures \ref{minpjet} and \ref{upub2} corresponds to a set of parameters $(B,\gmx,\lp,\delta)$ with 
$\delta \ge \dmin$, for which
the jet power is actually the minimum possible in the context of a hadronic model. 
In figure \ref{doppler} we plot the ratio $\delta/\dmin$ against the magnetic field strength. For the range $0.7 < B \le 10$ G where
the system is close to equipartition and the power of the jet is minimum, the ratio becomes larger than unity.

The conclusions drawn from the figures of the present section can be summarized as follows:
if one chooses to model the VHE $\gamma$-ray spectra using values of $(B,\gmx,\lp,\delta)$ 
that minimize both the required jet power and the total energy density of the system, i.e. model the system using
optimum energetic conditions, then one will be confronted with the requirement of a high Doppler factor.
On the other hand, acceptable fits of the TeV data using the minimum possible Doppler factor 
lead to a particle dominated system and to high values of $P_{\rm jet}^{\rm obs}$ for a wide range of $B$ values, i.e. the energetic
requirements in this case are higher. 

\section{Effects of other parameters}

In our numerical treatment we have kept fixed to specific values the power-law
index of the injected proton distribution, the radius of the emitting blob as well as the energy of the $\gamma$-ray photons $\epsilon_{\gamma}^{\rm obs}$.
Here we discuss possible effects of different adopted values on our results, since our initial assumption
could be critical for exploring the whole physically allowed parameter space for the 
system of 3C 279 at the observed state. 
\subsection{Power-law index $s$}
Throughout the present work we have assumed that the proton
distribution is a power law  of exponent $s=2$. In this case 
the energy per logarithmic interval is the same.
A different adopted value would 
affect the proton energy density quantitavely and therefore our energetics estimates. 
Let us for example consider a steep power law 
distribution with $s>2$. 
In this case more energy is injected to the protons with
the minimum Lorentz factor, whereas protons at the upper cutoff of the distribution,
which are responsible for the emitted radiation, carry only a small fraction
of the energy. 
Thus, for the same proton injection compactness the total number $N_{\rm p}$ and consequently the energy
density of protons $\up$ increases with increasing $s$. In addition,
the calculated $\gamma$-ray compactness for the same $\lp$ is lower. 
Thus, a larger proton compactness
$\lp$ and therefore energy density would be required in order to fit the TeV observations
with the same Doppler factor. 
As an extreme example, we considered a very steep proton distribution with 
the highest value for the power law exponent predicted by acceleration theory, i.e. $s=2.5$.
The fit to the TeV observations was obtained using: $B=40$ G, $\delta \approx 25$ and $\lp=10^{-1.3}$.
The proton energy density in this case is $\up=10^4$ erg/cm$^{-3}$, which is much higher than the values presented
in section 3.3. Therefore, the case of a flat proton distribution is rather conservative as far as the energetics is concerned.

\subsection{Radius of the emitting region}
 The fact that there is a very good agreement between the numerical results and
our analytical expressions, where the dependence on $R$ is explicit, makes possible the prediction of the effects
of a different adopted value on our results. Let us assume a more compact source with $R=3\times 10^{15}$ cm.
The quantities that are directly affected by a change in the source size are listed 
below in descending order in terms of their dependence on $R$:
\begin{enumerate}
\item
 The minimum Doppler factor for $B<\Bq$ (see equation \ref{dmin1}), which would be increased
by two orders of magnitude. Therefore, the `steep' branch of the plot in figure \ref{dmin-num} would be
shifted upwards by a factor of two in logarithmic scale. Note, that the constraint set by variability
arguments (relation \ref{tvar}) will not be violated since it implies an even lower limit to the Doppler factor
than previously. \\
\item
The equipartition magnetic field $\Beq$ given by equation (\ref{beq}), which for a certain $\delta$ will be increased
by a factor of 7. \\
\item
The magnetic field above which the feedback criterion is satisfied. We remind
that the scaling of $\Bq \propto R^{-3/5}$ has been derived after taking into account the constraint
of the Hillas criterion on $\gmx$.\\
\item
The minimum Doppler factor set by the automatic quenching is rather insensitive 
to changes of $R$ and it would be increased just by a factor of 2 (see equation (\ref{dmin2})).
\end{enumerate}
The proton energy density as well as the jet power calculated in the case of a blob moving with 
$\delta=\dmin$ will be also affected by a change in $R$. 
We remind that $\up \propto R^{-3} \delta^{-7/2}$ (see equation \ref{up}); the dependence on $R$ comes
through the constant $A$. Taking into account points (i) and (iv) above,
we find that $\up(\dmin) \propto R^4$ for $B< \Bq$ and $\up(\dmin) \propto R^{-4}$ for $B> \Bq$.
Since $\Bq$ also increases by a factor of 4, one expects to find significantly smaller values for $\up(\dmin)$
than those shown in figure \ref{densities} for a wider range of B-values (up to $\simeq 14$ G). 
The jet power obtained in this case has also a strong dependence on $R$, which is embedded in the constants
$\alpha_{1,2}$ and $b_{1,2}$ of equations (\ref{pjet1}) and (\ref{pjet2}) respectively. 
More specifically one finds that $\alpha_1 \propto R^{-2}$ and $\alpha_2 \propto R^2$ whereas 
$b_1 \propto R^{10/7}$ and $b_2 \propto R^{-4/7}$. The shape of the curve shown in figure \ref{jetpower}
would be transformed by shifting the different power-law segments vertically and horizontally, since 
the values of the magnetic field where the local minima occur would also be affected (see equations \ref{bhigh} and \ref{blow}).
It is important, however, to note that the minimum Doppler factor in this case would be extremely high (see also point (i) above).
This makes the scenario of a more compact $\gamma$-ray emitting region for 3C 279 less plausible.

\subsection{Energy of $\gamma$-ray photons}
\begin{figure}
 \centering
\resizebox{\hsize}{!}{\includegraphics{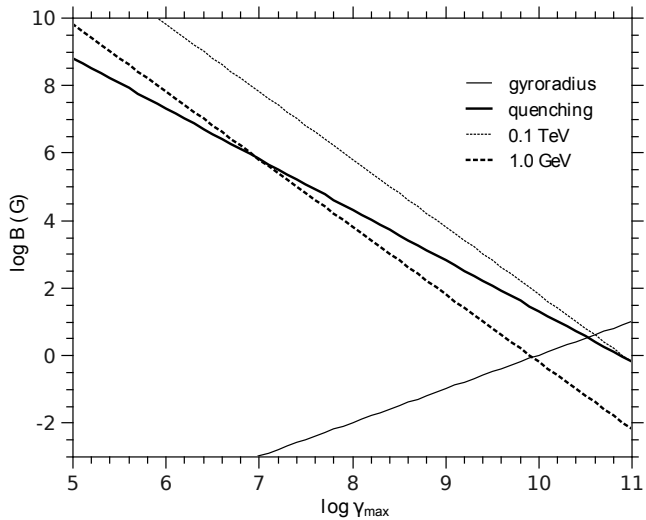}}
\caption{Log-log plot of characteristic values of the magnetic field against the maximum proton Lorentz factor.
The space below the thin solid line is not allowed, since it violates the Hillas criterion.
The feedback criterions is satisfied for values of $B$ and $\gmx$ that lie above the thick solid line, which represents $\Bq$. 
Finally, the thin and thick dotted lines $B$ and $\gmx$ values, that
correspond to $\epsilon_{\gamma}^{\rm obs}$ equal to 0.1 TeV and 1 GeV respectively.}
\label{bfield}
\end{figure}

Another question that naturally arises is how our results would change if the fitting method 
was used for  another set of $\gamma$- ray observations, e.g. in the GeV regime. Note that
some recent contemporary $\gamma$- and X-ray observations of 3C 279 (e.g. \citealt{abdo10, hayashida12})
could be an interesting case. In principle, the method outlined in sections 2 and 3 can be applied
with the only difference that the effects of quenching will not be seen, as automatic quenching cannot
set in for lower $\gamma$-ray energies at least for typical magnetic field strengths.
This is examplified in figure \ref{bfield}, where characteristic values of the magnetic field
are plotted against $\gmx$ in the extreme case of $\delta=1$. 
Specifically, the thick solid line corresponds to $\Bq$ (see equation (\ref{Bqgmx})) and divides
the parameter space into two regions. For $B$ values that lie above this line,
the feedback criterion is satisfied. 
The thin solid line corresponds to the Hillas criterion, i.e. $B \ge (\mpr c^2/eR) \gmx$.
Therefore, the  parameter space below this line is not allowed. 
Finally, the thin and thick dotted lines show the locus of $B$ and $\gmx$ values, that
correspond to $\epsilon_{\gamma}^{\rm obs}$ equal to 0.1 TeV and 1 GeV respectively, i.e.
$B=\epsilon_{\gamma}^{\rm obs} (\mpr c/\hbar e) \gmx^{-2} \delta^{-1}$ for $\delta=1$. 
Therefore, in the case where GeV  observations were used, one would need extremely
high values of the magnetic field, in order to see the effects of automatic quenching.
These would be even higher if one would take into account the exact value of the Doppler factor. 

\section{Discussion}
Hadronic models have been used extensively for fitting the $\gamma$-ray emission from Active Galactic Nuclei.
Usually, detailed fitting to the multiwavelength
spectrum of these objects requires, in addition to a population of relativistic protons,
 the presence of an extra leptonic component which is responsible
for the emission at lower energies (radio to UV or X-rays).

As it was shown recently \citep{SK07, PM11,petropoulou12}
compact $\gamma$-ray sources can be subject to photon quenching  which can result, if certain
conditions are met, to automatic $\gamma$-ray absorption and redistribution of the absorbed
$\gamma$-ray luminosity to electron-positron pairs and lower energy radiation. 
This is a non-linear loop which can operate even in the absence of an initial soft 
photon population and is expected to have direct consequences on the aforementioned models. 
For the present application other feedback loops, such as 
the Pair-Production-Synchrotron one \citep{mastkirk92}, 
 are less relevant as they usually operate 
at lower proton energies and at higher energy densities \citep{dimitrakoudisetal12}. 

Aim of the present paper is to set a general framework for investigating the effects of photon quenching on the parameter space
available for modelling $\gamma$-rays 
in the context of a hadronic model.
As an example we concentrated on the February 2006 observations of the blazar 3C 279 which were performed simultaneously 
 in the TeV and X-ray regimes. 
To this end we have focused only on fitting the high energy spectrum using a synchrotron
proton emission and treating conservatively the X-ray observations as an upper limit. We found that 
for a wide range of parameters, automatic photon quenching plays a crucial role as its onset
produces X-rays  which  violate the observations. To be able to assess its impact on the parameter space,
we have relaxed the usual goodness-of-fit method by accepting fits with $\chi_{\rm red}^2<1.5$. 
Our results indicate, in agreement with other researchers in the field,
that the hadronic model requires in general high magnetic fields.
Additionally we find that the presence of automatic quenching limits the proton luminosity (or, equivalently 
the proton energy density in the case of non substantial proton cooling) which, in turn, results in
a minimum value of the Doppler factor $\dmin$.  Interestingly enough, these ideas
do not apply to leptonic models because they favour much lower
values for the magnetic field. For these values photon quenching
does not operate, since its feedback criterion 
is not satisfied. The latter is a necessary condition for the emergence 
of automatic quenching and 
 requires for a given $\gamma$-ray energy 
a certain value of the magnetic field $\Bq$ , for the absoprtion loop to operate.
We have shown both 
analytically and numerically,
(see figures \ref{dmin-ana} and \ref{dmin-num} respectively), that
the minimum Doppler depends on the magnetic field strength in a different
way, depending on the relative relation of $B$ and $\Bq$. Specifically, 
if the magnetic field strength is 
above $\Bq$, then  $\dmin \propto B^{-1/7}$.
For values of the magnetic field with $ B < \Bq$ 
quenching is not relevant. However, we have shown using arguments based on the particles gyroradii,
that also in this case a lower limit to the Doppler factor exists, which  
depends strongly on $B$, i.e. $\dmin \propto B^{-3}$. 
Therefore if one wants to adopt a low
magnetic field for the radiating region, fits to the TeV $\gamma$-rays are still possible but at the
expense of a very high value of the Doppler factor.

The fact that quenching does not allow the Doppler factor to become smaller than some value
is intriguing and leads naturally to the investigation of the proton energy density inside
a blob which moves with this characteristic value.  In this case, 
we showed that there are two values of the magnetic field that minimize the energy content, one
corresponding to the $\delta$ branch with $B<\Bq$ and the other to the one with $B> \Bq$.
For values of the magnetic field between the two equipartition magnetic fields, 
the emission region is particle dominated
and the ratio $\up / \ub$ can be as large as $10^3$ (see figure \ref{densities}).
Note however, that if a Doppler factor twice the minimum one is adopted, then the calculated proton energy density
for the same magnetic field will be lower by almost an order of magnitude -- see equation (\ref{Lg}). 
Furthermore, we have calculated the jet power in the case 
where $\delta=\dmin$ and shown both analytically and numerically
that it is a function of $B$. The jet power is rather high
($10^{47} - 10^{49}$ erg/sec) for the whole range of $B$ values, as
expected in the context of hadronic modelling. 

We have repeated the above calculations by relaxing the condition
$\delta=\dmin$ while 
requiring the parameters to be such as to minimize 
the power of the jet. In this case we have shown that 
the energy content of the blob is also minimized (see figures \ref{minpjet} and \ref{upub2}).
For adopted $B$ values lying between the two equipartition values,
the jet power can be minimized, albeit at the cost of a high 
$\delta$ value. However, we have shown that
for magnetic field values outside of this
range, jet power minimization is not possible
 as this can be achieved only if $\delta<\dmin$. Thus,
 the existence of a minimum value for $\delta$
has indirect implications on the energetics of the system.

An interesting question is whether a detailed fitting to the X-ray
observations of 3C~279
with the addition of an extra leptonic component
would bring any change to the basic ideas presented here. 
In this case apart from automatic quenching, the linear absorption of $\gamma$-rays
on the X-ray photons emitted by the leptonic component, is also at work. 
However, by including this component and
repeating our numerical calculations of \S3,
we found that our results do not change.  
This is due to the fact that the X-ray luminosity of 3C~279 
is rather low and therefore the effects of linear 
$\gamma\gamma$ absorption are minimal, at least up to the 
compactnesses above which the automatic photon quenching sets in. 
Apart from the absorption of $\gamma$-rays on the 
synchrotron photons emitted by the `extra' leptonic component discussed above, 
inverse compton scattering of these photons to higher energies by the same
leptonic component would be an additional mechanism at work.
The upscattered photons would lie in the hard X-ray and $\gamma$-ray energy range and would
affect our calculations only if their luminosity $L_{\rm ssc}$ would be comparable to 
that carried by the synchrotron component $L_{\rm syn}$.
While estimating the ratio $u_{\rm  syn}/ \ub$ for a wide range of $B$ and $\dmin$ values
used in the present work (see figure \ref{dmin-num}) we find that this
is always much smaller than unity, and therefore the
inverse Compton scattering of an extra leptonic component
which would fit the radio to X-ray observations does not
interfere to the quenching mechanism. 

The estimation method proposed in the present work
can be regarded as an extension of the widely used method for estimating the equipartition
magnetic field using radio observations. In our case, the leptonic synchrotron component is replaced by the
proton synchrotron emission and the radio by the VHE $\gamma$-ray observations. 
The innovative feature of our method is the estimation
of a minimum Doppler factor, which is the result of
automatic photon quenching. This can be more of relevance
to TeV observations rather
than the Fermi ones as quenching at GeV energies requires
very high values of the magnetic field -- see figure \ref{bfield}. 
The fact that our numerical results are in very good agreement
with the analytical calculations offers a fast yet robust way for estimating
many physical quantities of the $\gamma$-ray emitting region of TeV blazar jets.
We caution however the reader, that the effects of automatic quenching, 
i.e. when no soft photons are initially present in the 
emitting region, can be seen only in a self-consistent treatment of the radiative transfer problem.

%
%
%
\section{aknowledgements}
This research has been co-financed by the European Union 
(European Social Fund -- ESF) and Greek national funds through the Operational Program 
"Education and Lifelong Learning" of the National Strategic Reference Framework 
(NSRF) - Research Funding Program: Heracleitus II. Investing in knowledge society 
through the European Social Fund. We would like to thank Dr. Anita Reimer and the
anonymous referee for usefull comments on the manuscript.

\bibliographystyle{mn2e}
\bibliography{0417.bib}

\label{lastpage}
\end{document}